\documentclass[fleqn,usenatbib]{mnras}
\usepackage{newtxtext,newtxmath}
\usepackage[T1]{fontenc}
\usepackage{ae,aecompl}

\usepackage{graphicx}	% Including figure files
\usepackage{amsmath}	% Advanced maths commands
\usepackage{amssymb}	% Extra maths symbols

\usepackage{color}

\newcommand{\phpl}{Phys.\ Plasmas}
\newcommand{\soph}{Sol.\ Phys.}
\newcommand{\newa}{New Astron.}
\newcommand{\astronach}{Astron.\ Nach.}

\title{Numerical Simulation of Helical Jets at Active Region Peripheries}

\author[P. F. Wyper et al.]{
Peter F. Wyper,$^{1}$\thanks{E-mail: peter.f.wyper@durham.ac.uk (PFW)}
C. Richard DeVore,$^{2}$
Spiro K. Antiochos$^{2}$
\\
$^{1}$Department of Mathematical Sciences, Durham University, Durham, DH1 3LE, UK\\
$^{2}$Heliophysics Science Division, NASA Goddard Space Flight Center, 8800 Greenbelt Road, Greenbelt MD 20771, USA
}

\date{Accepted XXX. Received YYY; in original form ZZZ}

\pubyear{2019}

\begin{document}
\label{firstpage}
\pagerange{\pageref{firstpage}--\pageref{lastpage}}
\maketitle

%This is a simple template for authors to write new MNRAS papers.
%The abstract should briefly describe the aims, methods, and main results of the paper.
%It should be a single paragraph not more than 250 words (200 words for Letters).
%No references should appear in the abstract.
\begin{abstract}
Coronal jets are observed above minority polarity intrusions throughout the solar corona. Some of the most energetic occur on the periphery of active regions where the magnetic field is strongly inclined. These jets exhibit a non-radial propagation in the low corona as they follow the inclined field, and often have a broad, helical shape. We present a three-dimensional magnetohydrodynamic simulation of such an active-region-periphery helical jet. We consider an initially potential field with a bipolar flux distribution embedded in a highly inclined magnetic field, representative of the field nearby an active region. The flux of the minority polarity sits below a bald-patch separatrix initially. Surface motions are used to inject free energy into the closed field beneath the separatrix, forming a sigmoidal flux rope which eventually erupts producing a helical jet. We find that a null point replaces the bald patch early in the evolution and that the eruption results from a combination of magnetic breakout and an ideal kinking of the erupting flux rope. We discuss how the two mechanisms are coupled, and compare our results with previous simulations of coronal-hole jets. This comparison supports the hypothesis that the generic mechanism for all coronal jets is a coupling between breakout reconnection and an ideal instability. We further show that our results are in good qualitative and quantitative agreement with observations of active-region periphery jets.
\end{abstract}

\begin{keywords}
Sun: corona -- Sun: filaments, prominences -- Sun: flares -- Sun: magnetic fields -- magnetic reconnection
\end{keywords}

%%%%%%%%%%%%%%%%%%%%%%%%%%%%%%%%%%%%%%%%%%%%%%%%%%

%%%%%%%%%%%%%%%%% BODY OF PAPER %%%%%%%%%%%%%%%%%%

\section{Introduction}
Coronal jets are impulsive, collimated ejections of plasma that originate low in the solar atmosphere and propagate outwards along the ambient magnetic field. They are observed most readily in extreme ultraviolet (EUV) and X-rays \citep[e.g.,][]{Shimojo1996,Savcheva2007,Cirtain2007,Nistico2009}, originate above minority polarity intrusions \citep[e.g.,][]{Shimojo1998}, and generally involve the impulsive onset of reconnection between the field closing locally to the minority polarity and the surrounding open, or distantly closing, magnetic field. Coronal jets have various morphologies \citep{Nistico2009,Moore2010}: some form a tapered narrow spire of hot plasma, whilst others form broad, helical spires containing both cool and hot components (relative to the ambient corona). For recent reviews of jets see \citet{Innes2016} and \citet{Raouafi2016}.

\begin{figure*}
\includegraphics[width=0.95\textwidth]{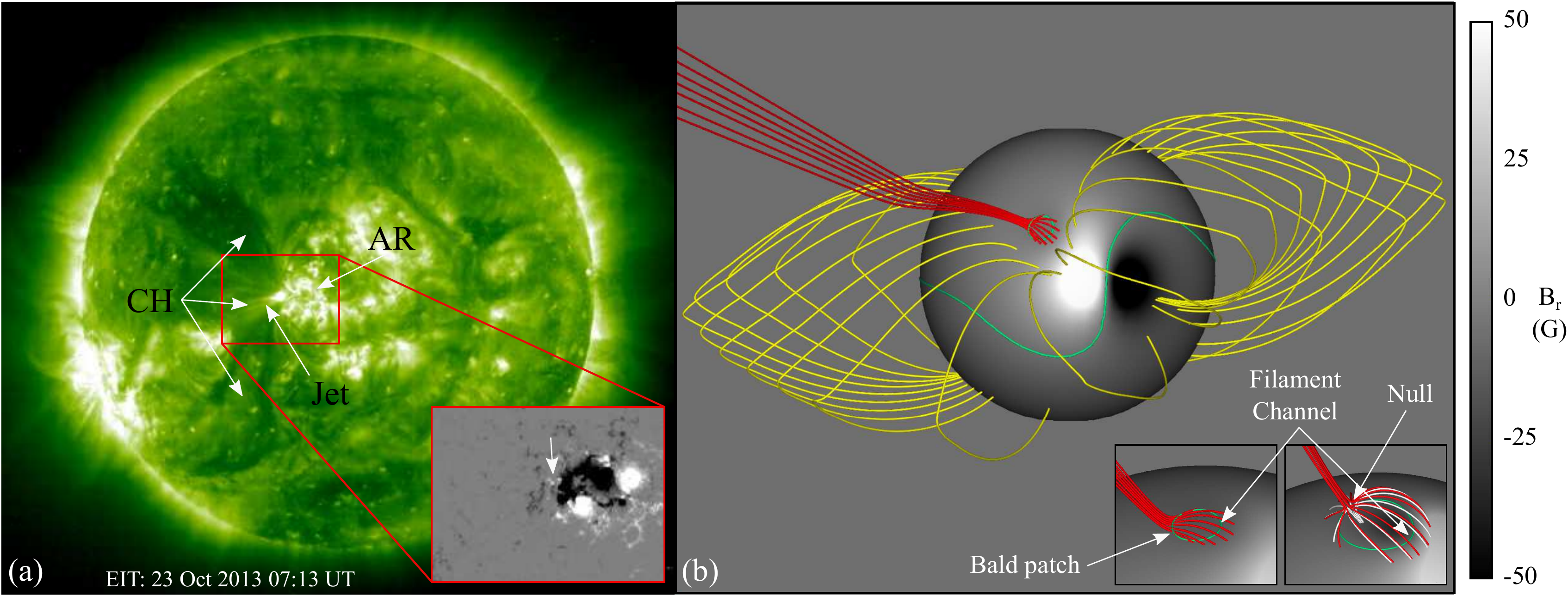}
\caption{(a) An example of an ARP jet that escapes along the open field lines of a coronal hole \citep[see][for more details]{Chandra2017}. AR = active region; CH = coronal hole. The inset shows the surface magnetic field: the jet originates from the highlighted minority polarities east of the active region. (b) A model potential field containing a bald patch in an open-field region adjacent to a strong active region. Red field lines pass through the bald patch, and yellow field lines show the helmet-streamer boundary. The green contour shows the PIL. Left inset: close-up view of the bald patch. Right inset: field lines near a coronal null in a field with a slightly stronger minority-polarity patch.}
\label{fig:obs}
\end{figure*}

Due to the magnetic field strengths and magnetic fluxes involved, the largest and most powerful jets usually occur in the vicinity of active regions and are associated with opposite-polarity satellite spots. Those originating near the centres of active regions are confined along closed coronal loops that guide the jet and its associated accelerated particles along a curved path back to the surface \citep[e.g.,][]{Hanaoka1996,Yang2012,Cheung2015,Li2017,Li2017b}. Depending upon the relative size of the jet region versus the loop length, these events are sometimes also classified as confined eruptions or flares \citep[e.g.,][]{Sun2013}. Jets originating at the edges of active regions, on the other hand, form at the base of much longer loops. The field lines of these loops are highly inclined away from the vertical at the solar surface. Such active-region-periphery (ARP) jets follow this highly inclined field in their early stages \citep[e.g.,][]{Canfield1996,Guo2013,Hong2016,Mulay2016}. These jets often have a helical morphology, and contain both a hot, tenuous and a cooler, denser plasma component \citep[relative to the surrounding corona; e.g.,][]{Mulay2017}. The cool component is sometimes referred to as a surge \citep[e.g.,][]{Canfield1996}. Additionally, depending upon the global topology of the coronal magnetic field, ARP jets are occasionally launched into the open field of a low-latitude coronal hole \citep[e.g.,][]{Mulay2016,Chandra2017}. In these cases, the field along which the jet propagates is highly inclined near the solar surface and transitions to approximately radial further out. Figure \ref{fig:obs}(a) shows an example of such an ARP jet propagating into a low-latitude coronal hole. These ARP jets can produce sizeable jet-like coronal mass ejections (CMEs) \citep[e.g.,][]{Wang2002} and are associated with impulsive solar energetic particle (SEP) events \citep[e.g.][]{Nitta2015,Innes2016,Bucik2018,Glesener2018}. 

Magnetic extrapolations of the field at the base of ARP jets find one of two magnetic topologies. In some cases the minority polarity is separated from the locally open surrounding field by the fan plane of a three-dimensional null point in the same manner as coronal-hole jets \citep[e.g.,][]{Mandrini2014,Zhu2017}. {However, unlike coronal-hole jets, the ARP-jet null point resides off to one side of the separatrix, near the solar surface.} In other cases the separatrix is formed by field lines that skiff the surface at a so-called bald patch \citep[e.g.,][]{Schmieder2013,Chandra2017}. Figure \ref{fig:obs}(b) shows a potential-field source-surface (PFSS) model demonstrating the field structure of both topologies. The PFSS is constructed with a large active region at the equator that forms a low-latitude corona hole \citep{Antiochos2011,Titov2011}. A small minority polarity was placed next to the active region within the coronal hole, and changes to its size and strength raise or lower the null relative to the surface. As the local field is highly inclined, the coronal null sits below the apex of the separatrix so that field lines directly above the null are dipped. These dipped field lines form the bald patch when the null sinks below the surface. This shows the natural link between the two topologies and why a high local field inclination is needed to produce a bald patch in a coronal hole \citep[see also][]{Titov1993,Bungey1996,Muller2008}. In both cases, the open field lines that touch the separatrix are angled away from the active region near the surface before bending upwards to become radial. 

Observations show that the minority polarities at the bases of ARP jets are constantly evolving, injecting free energy into the field beneath the separatrix through a combination of flux emergence, flux cancellation, and relative motion \citep[e.g.,][]{Yan2012,Mulay2016}. Observations and extrapolations have also revealed that like some coronal-hole jets \citep[][]{Sterling2015}, some ARP jets are generated when a mini-filament erupts \citep[][]{Mandrini2014,Hong2016,Sterling2016,Zhu2017}. The mini-filament usually forms along a section of the polarity inversion line (PIL) that separates the minority polarity from the strongest field at the edge of the active region, {i.e., along the section of PIL highlighted in Figure \ref{fig:obs}(b). Therefore, in general the strapping field above the filament channel in ARP jets is aligned with the overlying background field. Such a configuration makes external reconnection and removal of the strapping field difficult to achieve. It also suppresses instabilities such as torus, for example. How, then, do such filament channels erupt and transfer their twist to form helical ARP jets?} 

Previous dynamic models of active-region jets have focussed on jets confined along the relatively short coronal loops rooted near the centres of active regions \citep[e.g.,][]{Gontikakis2009,Torok2009,Archontis2010,Cheung2015,Wyper2016,Wyper2016b}. {Other jet studies have examined how flux emergence forms filament channels that erupt to produce helical jets in an inclined uniform background field \citep[e.g.][]{Archontis2013,Moreno-Insertis2013, Fang2014}. However, the ambient field direction in those experiments is opposite to the strapping field above the filament channel created by the emergence, thereby readily allowing external reconnection and the ultimate eruption of the filament channel to occur. In this work, we focus on a configuration more typical of ARP jets, in which external reconnection is able to occur much less readily}. In particular, we present a magnetohydrodynamic simulation where surface motions form a filament channel beneath a bald-patch separatrix embedded in a highly inclined ambient field {aligned with the strapping field above the filament channel}. This simulation is an extension of our model for coronal-hole jets \citep{Wyper2017,Wyper2018}, in which a filament channel is formed beneath the spine-fan topology of a three-dimensional (3D) null point surrounded by nearly vertical ambient field. Breakout reconnection at the null leads to the eruption of the filament channel, producing a helical jet. We find that, {despite the challenges presented by the bald-patch topology and} like the coronal-hole jet model, breakout reconnection is integral to the eruption of the filament channel in this simulated ARP jet. However, we also find that the kink instability plays an important role in the late stages of the breakout process and jet generation. The simulation setup is described in \S \ref{sec:setup}, \S \ref{sec:results} describes our results, and \S\S \ref{sec:discussion} and \ref{sec:summary} discuss and summarise our findings. 

\begin{figure}
\includegraphics[width=\columnwidth]{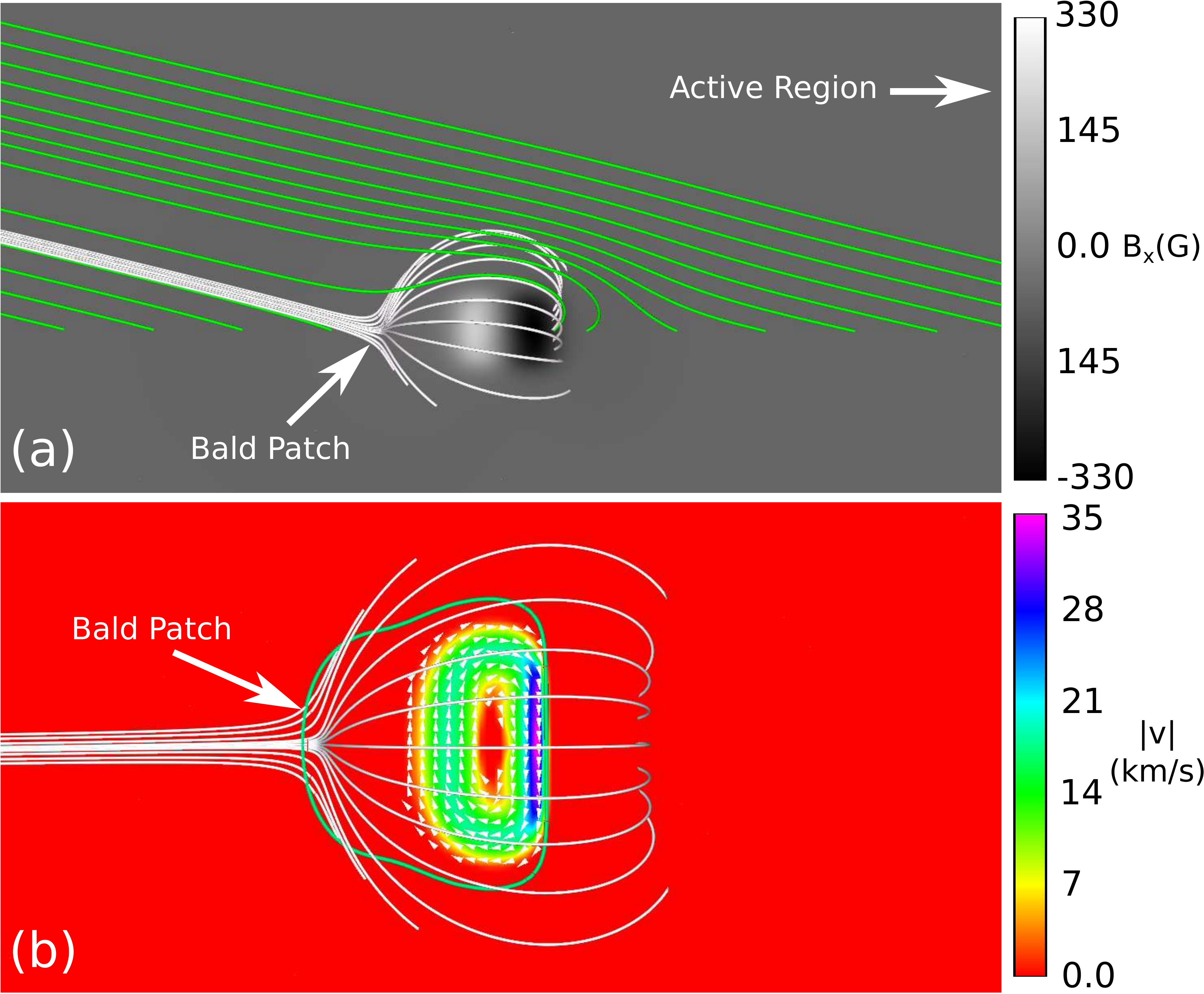}
\caption{Simulation setup. (a) The magnetic field. Greyscale shading shows $B_{x}$. Silver field lines pass through the bald patch and show the separatrix surface. (b) Surface driving flows. Colour denotes velocity magnitude. The PIL is shown in green.}
\label{fig:setup}
\end{figure}

\begin{figure*}
	\includegraphics[width=2\columnwidth]{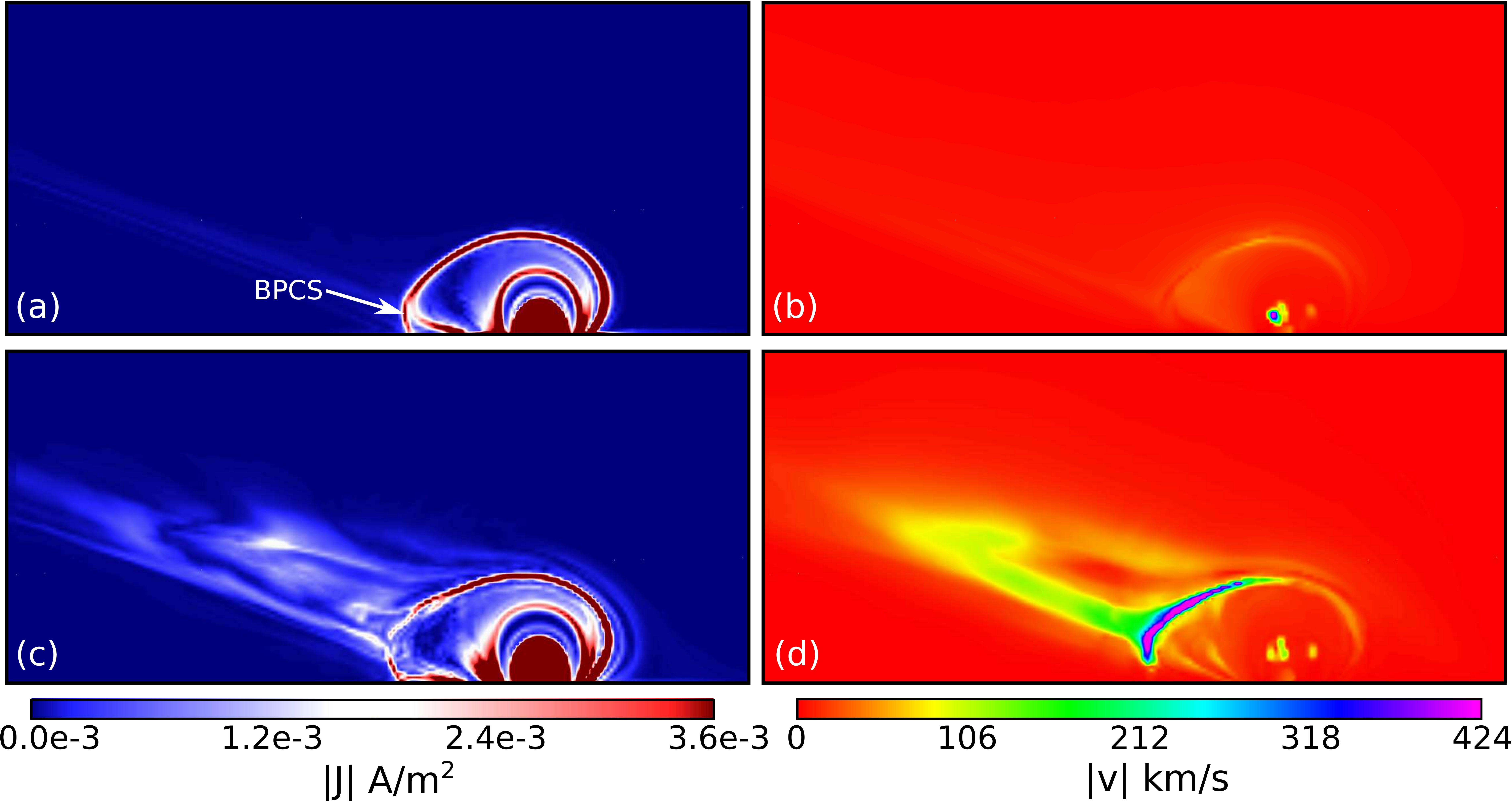}
   \caption{External reconnection just before (a,b; $t \approx 10$\,min $30$\,s) and soon after (c,d; $t \approx 14$\,min) initiation. Left: current density. Right: velocity magnitude. Note that both colour scales are saturated. BPCS = bald-patch current sheet.}
   \label{fig:bp}
\end{figure*}
%t=180, 240

\section{Simulation Setup}
\label{sec:setup}
We start with a potential magnetic bipole embedded in a uniform, inclined magnetic field, as shown at $t=0$ in Figure \ref{fig:setup}(a). The uniform background field represents to lowest order a nearby active region, positioned to the right of the bipole in the figure as indicated. This background field is inclined at angle $\theta = 70^{\circ}$ counter-clockwise from the vertical. The embedded bipole is constructed from multiple sub-surface dipoles in the manner of \citet{Wyper2018}. The resulting total magnetic field contains a bald patch along the section of PIL to the left of the bipole. Field lines touching the bald patch (shown in silver in Fig.\ \ref{fig:setup}) form a separatrix between the regions of open and closed flux. The rest of the setup is the same as our previous investigations: the domain is a closed Cartesian box; the plasma is uniform; gravity, stratification, and plasma heating are neglected; and the ideal magnetohydrodynamic equations are solved with an adiabatic energy equation. Consequently, reconnection occurs through numerical diffusion, and changes to temperature and density arise purely from compression/expansion. The lower boundary is closed and line-tied, whilst the side boundaries are open but placed sufficiently far away that the main jet disturbance does not reach them before the simulation is halted. The grid is refined adaptively, based upon local gradients in the magnetic field \citep{Karpen2012}. We use four levels of grid refinement in this simulation. 

Free energy is introduced by tangential surface motions that follow the contours of $B_{x}$, the normal component of the surface magnetic field. The driving is ramped up smoothly, held at a constant speed, and then ramped down again. The spatial profile of the flow is shown in Figure \ref{fig:setup}(b); the PIL is the green curve. This flow adds a broad twist to the field beneath the separatrix, with the strongest shearing concentrated along the PIL at the centre of the bipole. As in our previous simulations, this creates a filament channel at that location. 

The equations were solved in non-dimensional units, which can be scaled to solar values through choices of representative scales for length ($L_{s}$), mass density ($\rho_{s}$), and magnetic field strength ($B_{s}$). Here we choose $L_{s} = 5$\,Mm, $\rho_{s} = 2\times 10^{-14}$\,g\,cm$^{-3}$, and $B_{s} = 20$\,G to scale our results to solar active-region jets. These values are used throughout, but we note that our results can be rescaled to a particular event by multiplying these typical values by an appropriate factor; see \citet{Wyper2018} for details. With these scalings, the ambient plasma temperature, density, and pressure are $1.2\times 10^{6}$\,K, $2\times 10^{-14}$\,g\,cm$^{-3}$, and $4$\,dyn\,cm$^{-2}$, respectively.  The background uniform magnetic field has strength $21.5$\,G and the peak field strength in the minority polarity is $338$\,G {\citep[cf.\ ][]{Schmieder2013,Zhu2017}}, so the plasma $\beta$ $\approx 2\times 10^{-1}$ and $\approx 9\times 10^{-4}$, respectively. The maximum width of the separatrix dome is $\approx 30$\,Mm, whilst the minimum grid spacing is $\approx 260$\,km. The maximum driving speed is $|V| \approx 35$\,km\,s$^{-1}$, about $0.5$\% of the local Alfv\'{e}n speed and  $20$\% of the sound speed. Thus, the free energy builds up quasi-statically in the closed magnetic field. The driving is ramped up over  $3$\,min at the start of the simulation, held constant for $21$\,min, and then ramped down to zero over $3$\,min. For details of the driving profile, equations solved, and boundary conditions see \citet{Wyper2018}. 

\begin{figure}
\includegraphics[width=\columnwidth]{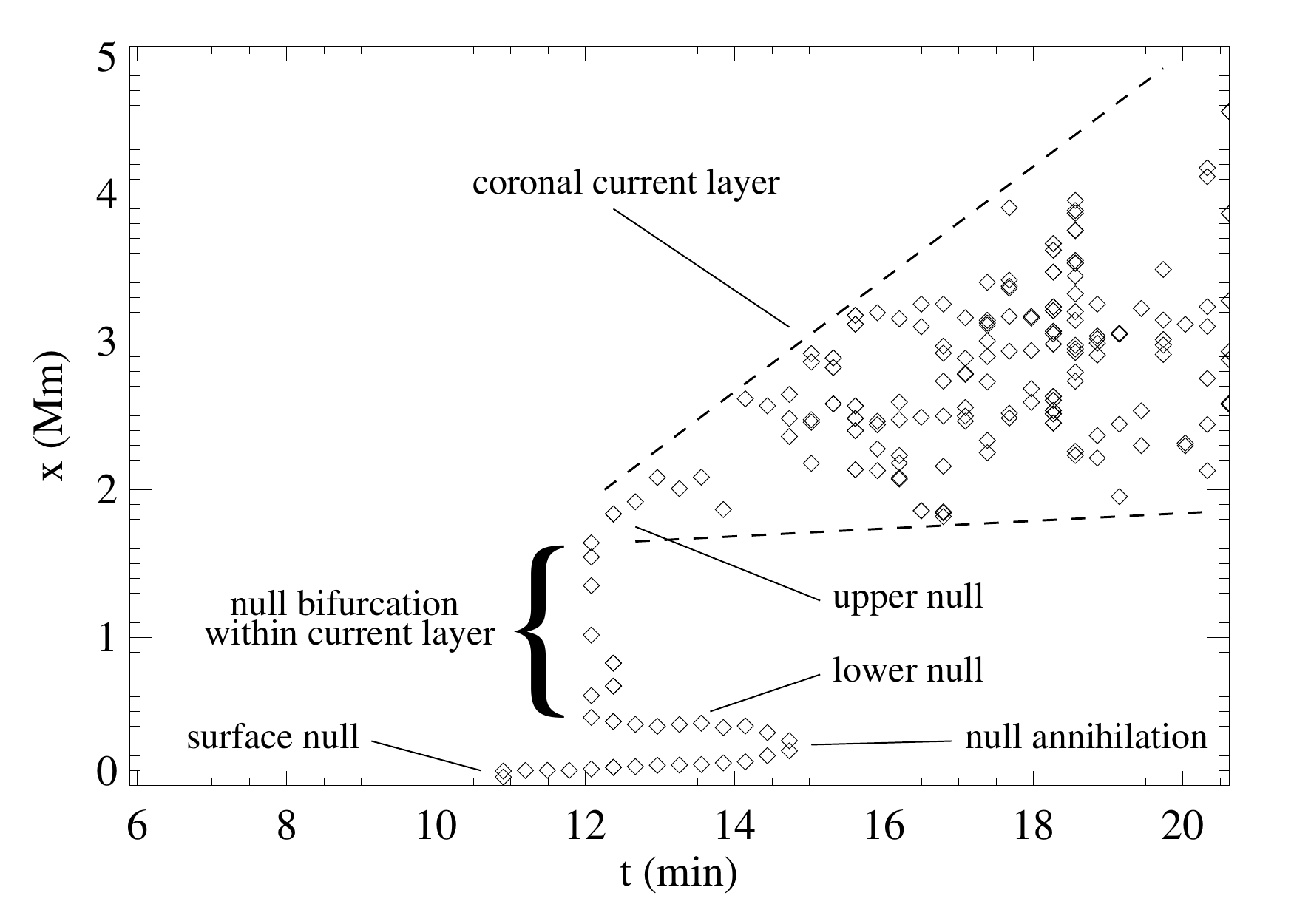}
\caption{The heights of identified nulls around the time of reconnection initiation in the bald-patch current sheet (BPCS).}
\label{fig:null}
\end{figure}

\begin{figure}
\includegraphics[width=0.9\columnwidth]{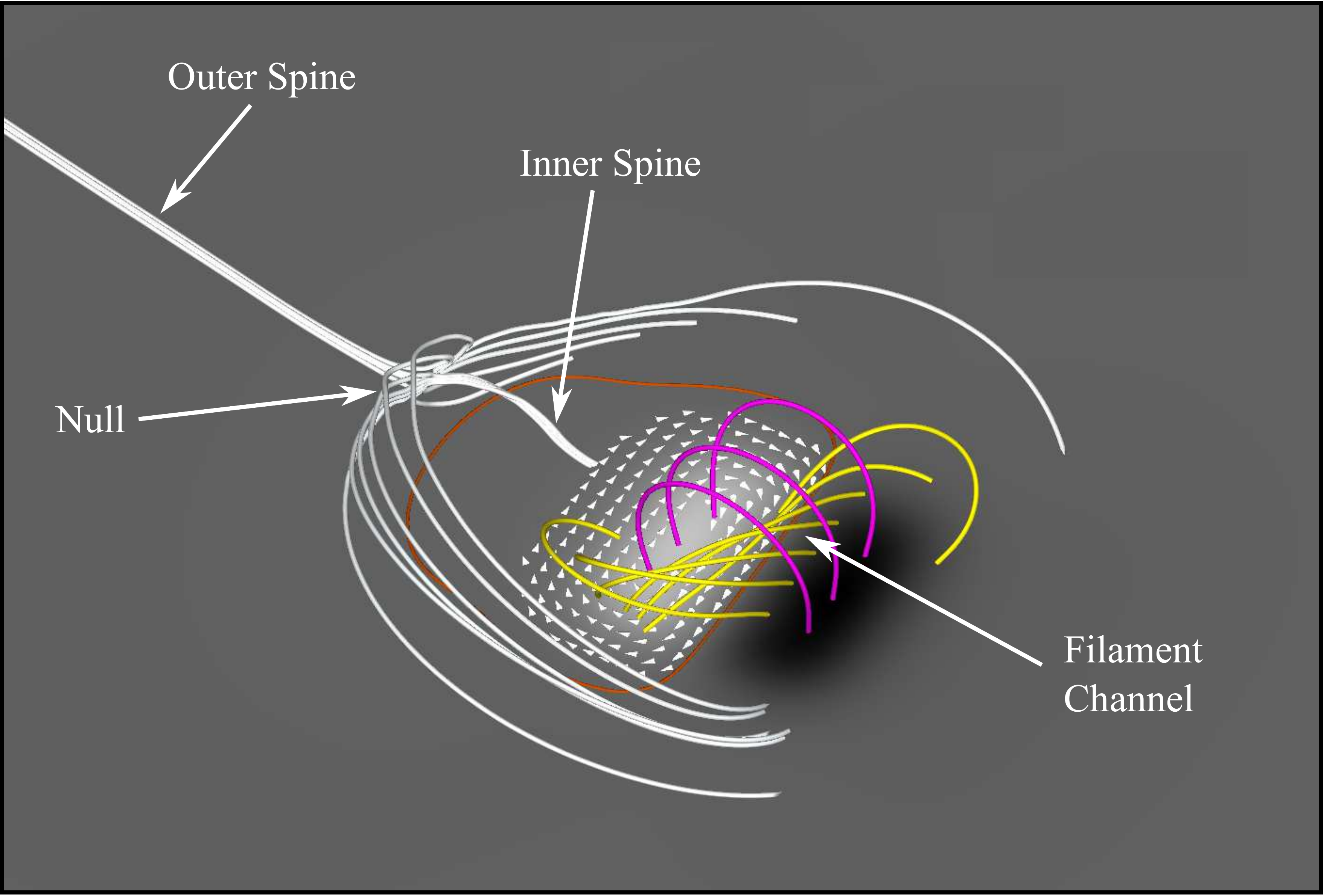}
\caption{Upper null topology after initiation of external reconnection ($t \approx 14$ min). Surface shading as in Fig.\ \ref{fig:setup}. The orange contour shows the PIL and white arrows show the surface driving. }
\label{fig:null2}
\end{figure}

\section{Results}
\label{sec:results}
\subsection{Initiation of External Reconnection}
\label{sec:recon}
Once the driving begins, closed field lines connecting to the minority polarity are sheared, storing free magnetic energy beneath the separatrix. The magnetic pressure increase expands the separatrix vertically but also laterally, so that it extends over the bald patch. This lateral expansion bends the field lines threading the bald patch back onto themselves, forming the current sheet (BPCS) shown in Figure \ref{fig:bp}(a) {within the overarching separatrix-surface current layer. Emanating from the BPCS into the closed-field region, highly inclined to the solar surface, is a current sheet associated with the outer edge of the driving region. Also visible are two other strong current regions: the innermost volumetric current formed within the filament channel, and an arching current layer above this channel and its overlying arcade of loops. The filament-channel volume, the arching layer and the highly inclined layer currents are generated directly by the shearing footpoint motions shown in Figure \ref{fig:setup}.} {Negligible reconnection occurs within all of these internal current structures during the flux-rope formation and expansion phase of the evolution. The bald-patch current sheet has not yet began to reconnect at the time shown in Figure \ref{fig:bp}(a,b), so there is negligible plasma flow near the bald patch.} 

The formation of 3D null points in the current sheet marks the onset of reconnection. Using the tri-linear method of \citet{Haynes2007} \citep[see][for details of the implementation]{Wyper2016b}, the null points in the simulation volume were identified. Figure \ref{fig:null} shows the heights of the identified nulls versus time as the reconnection begins. Around $t = 11$\,min a (surface) null forms on the simulation boundary. As discussed in \S \ref{sec:discussion}, this null is an artifact of the surface boundary conditions, and its formation destroys the bald patch beneath it. Soon after ($t \approx 12$\,min), a group of nulls briefly appears within the current sheet before annihilating to leave just two nulls (upper and lower). The lower null moves down and annihilates with the surface null, whilst the upper null climbs higher into the corona and spawns multiple additional nulls within the lengthening current sheet through further bifurcations. Field lines outlining the spine-fan structure of the upper null are shown in Figure \ref{fig:null2}. {The ambient field's strong inclination from the vertical direction and alignment with the strapping field above the filament channel ensure that once the upper null forms, along with the new nulls spawned from it, it remains relatively close to the surface and above the section of the PIL where it formed.} 
The upper null facilitates interchange reconnection of sheared closed field and unsheared open field within the current sheet, initiating the plasma flow shown in Figure \ref{fig:bp}(d). Moreover, the formation of the upper and lower nulls marks a change in global topology, wherein the fan plane of the upper null becomes the open-closed separatrix. This evolution will be discussed further in \S \ref{sec:discussion}. 

\begin{figure*}
	\includegraphics[width=2\columnwidth]{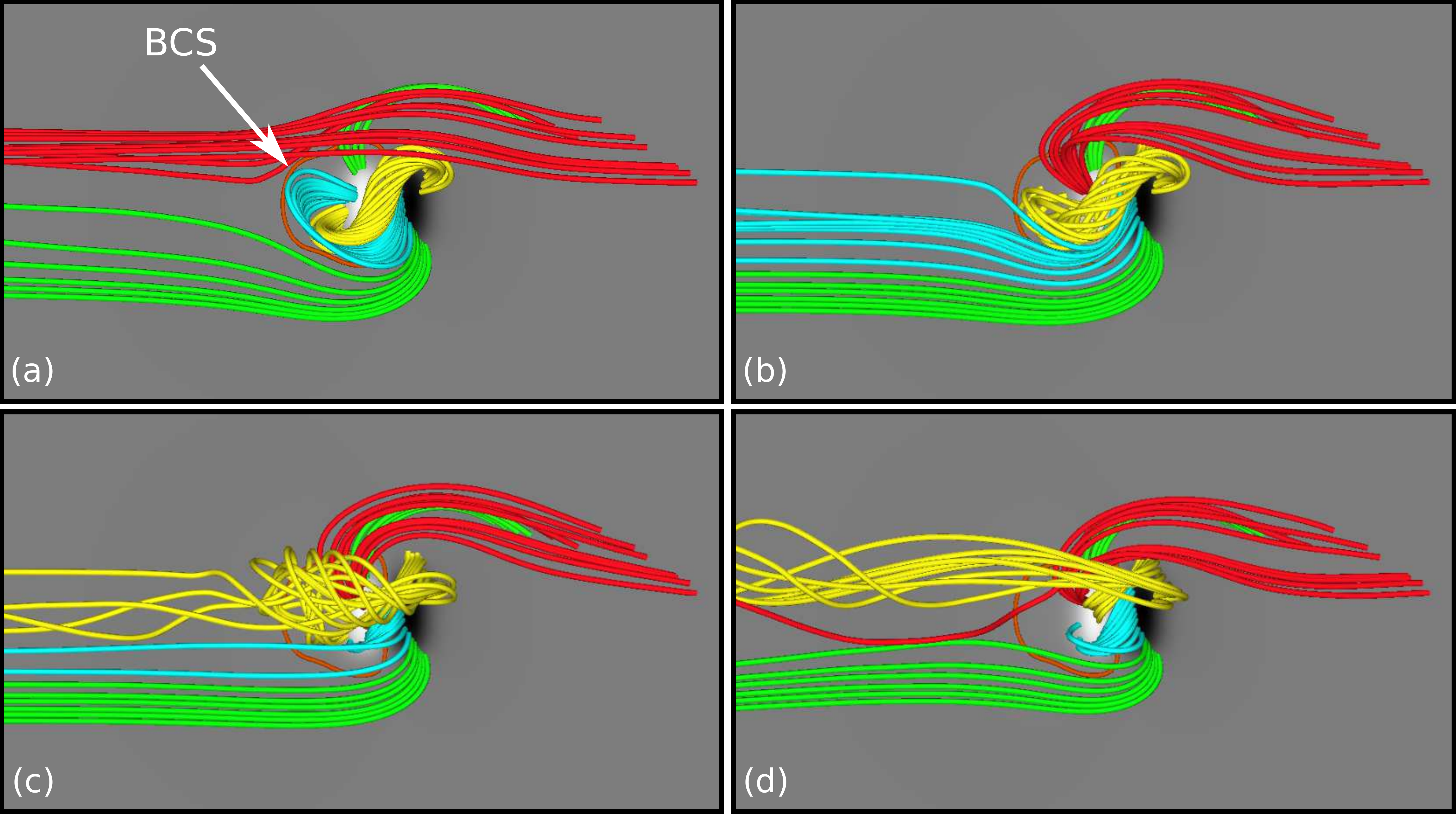}
  \caption{Top view of field lines showing the breakout evolution (a,b), flux-rope eruption (c), and jet (d). (a) $t \approx 26$\,min $30$\,s. (b) $t \approx 47$\,min. (c) $t \approx 51$\,min $30$\,s. (d) $t \approx 55$\,min. Surface shading as in Fig.\ \ref{fig:setup}. The PIL is shown in orange. BCS = breakout current sheet. An animation is available online.}
  \label{fig:flstop}
\end{figure*}
%t = 45, 80, 87, 93

\begin{figure}
	\includegraphics[width=\columnwidth]{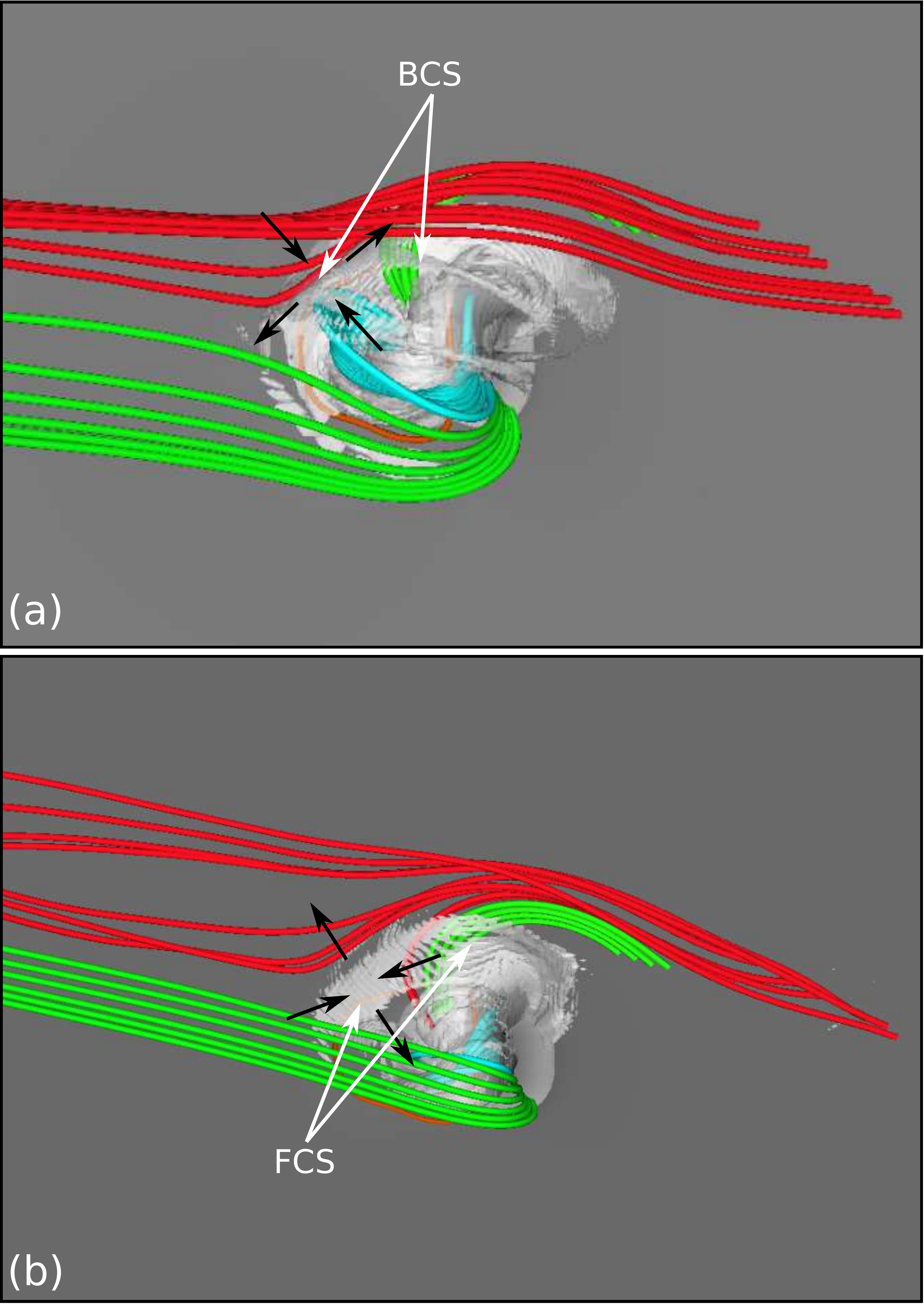}
  \caption{Isosurfaces of current density ($|J| = 3.6\times10^{-3}$A m$^{-2}$) showing (a) the breakout current sheet (BCS) and (b) the flare current sheet (FCS). (a) $t \approx 26$\,min $30$\,s. (b) $t \approx 59$\,min. Field lines as in Fig.\ \ref{fig:flstop}. The yellow filament channel/flux-rope field lines have been omitted to better see the current sheets.}
  \label{fig:current}
\end{figure}
%|J| = 0.3, t = 45, 100

\subsection{Helical Jet}
Quasi-steady reconnection outflows continue from the upper null whilst the driving is maintained. The surface driving motions are halted at $t \approx 26$\,min $30$\,s. By this time a sigmoidal flux rope has developed (Fig.\ \ref{fig:flstop}(a), yellow field lines). The flux rope forms as a result of the sharp gradient in the driving profile near the centre of the bipole. This gradient forms a current sheet near the surface that converts the shear within the filament channel to twist via slow closed-closed \citep[tether cutting;][]{Moore2001} reconnection. The surrounding null-point topology at this time is shown by the other sets of field lines, forming four flux regions. Field lines from a sub-region of the strapping field are shown in cyan; they constrain one end of the sigmoid from expanding laterally, rather than the centre from expanding upwards. It is the removal of these side field lines that leads to the eruption in our simulation. Green field lines show open and closed side-lobe regions on either side of the strapping field. Red field lines show the open field that is directly opposite the strapping field, across the null.

A current sheet resides on the open-closed boundary between the cyan strapping field and the red overlying field. Breakout reconnection within this sheet (marked BCS in the figure) removes the strapping field by reconnecting it onto the green side-lobe field. An isosurface of current showing the breakout current sheet is shown in Figure \ref{fig:current}(a), along with arrows indicating the direction of reconnection inflows and outflows. Over the next $20$ minutes or so, breakout reconnection slowly removes the cyan field lines until all of the strapping field has opened and the corresponding red overlying field with which it reconnects has closed (Fig.\ \ref{fig:flstop}(b); see also the animation). Concurrently, further turns of twist develop in the flux rope through closed-closed {tether-cutting} reconnection in the current sheet beneath it. This closed-closed reconnection follows from the expansion and rotation of the flux rope towards the breakout current sheet. Recall that, throughout this slow evolution, no driving is applied. {Therefore, the closed-closed reconnection is not due to the surface shearing directly; rather, it arises from the internal relaxation of the structure as the previously injected shear is converted into twist.} The evolution here is physically very similar to that seen by \citet{Lynch2009} in CME simulations, where the CME flux rope rotated due to the twist that developed from flare reconnection below the ejection. As the flux rope in our simulation gets closer to the breakout current sheet, it rotates further until it is at an angle $\approx 90^{\circ}$ relative to the section of PIL along which it formed at the centre of the bipole (Fig.\ \ref{fig:flstop}(c)). At around this time, the flux rope begins to reconnect onto open field lines within the breakout current sheet, transferring its twist. Once this reconnection of the flux rope begins, its field lines open rapidly (Fig.\ \ref{fig:flstop}(d)). Note that only the end of the flux rope rooted in the negative (majority) polarity opens, as shown in Figure \ref{fig:flstop}(d). The end rooted in the positive (minority) polarity remains closed and reforms the filament channel, but with reduced shear (not shown). 

\begin{figure*}
	\includegraphics[width=2\columnwidth]{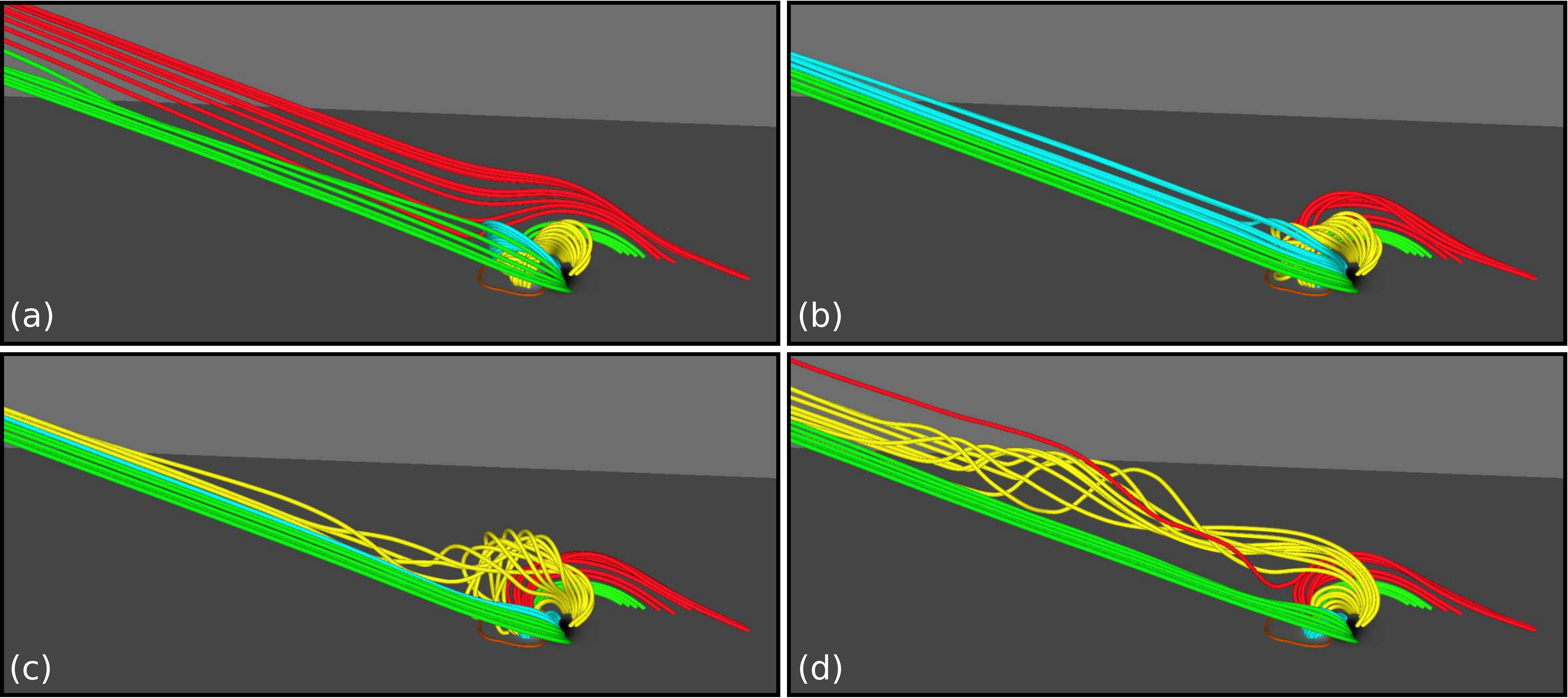}
  \caption{Side view of field-line evolution shown in Fig.\ \ref{fig:flstop}. An animation is available online.}
  \label{fig:flsangle}
\end{figure*}

\begin{figure*}
	\includegraphics[width=2\columnwidth]{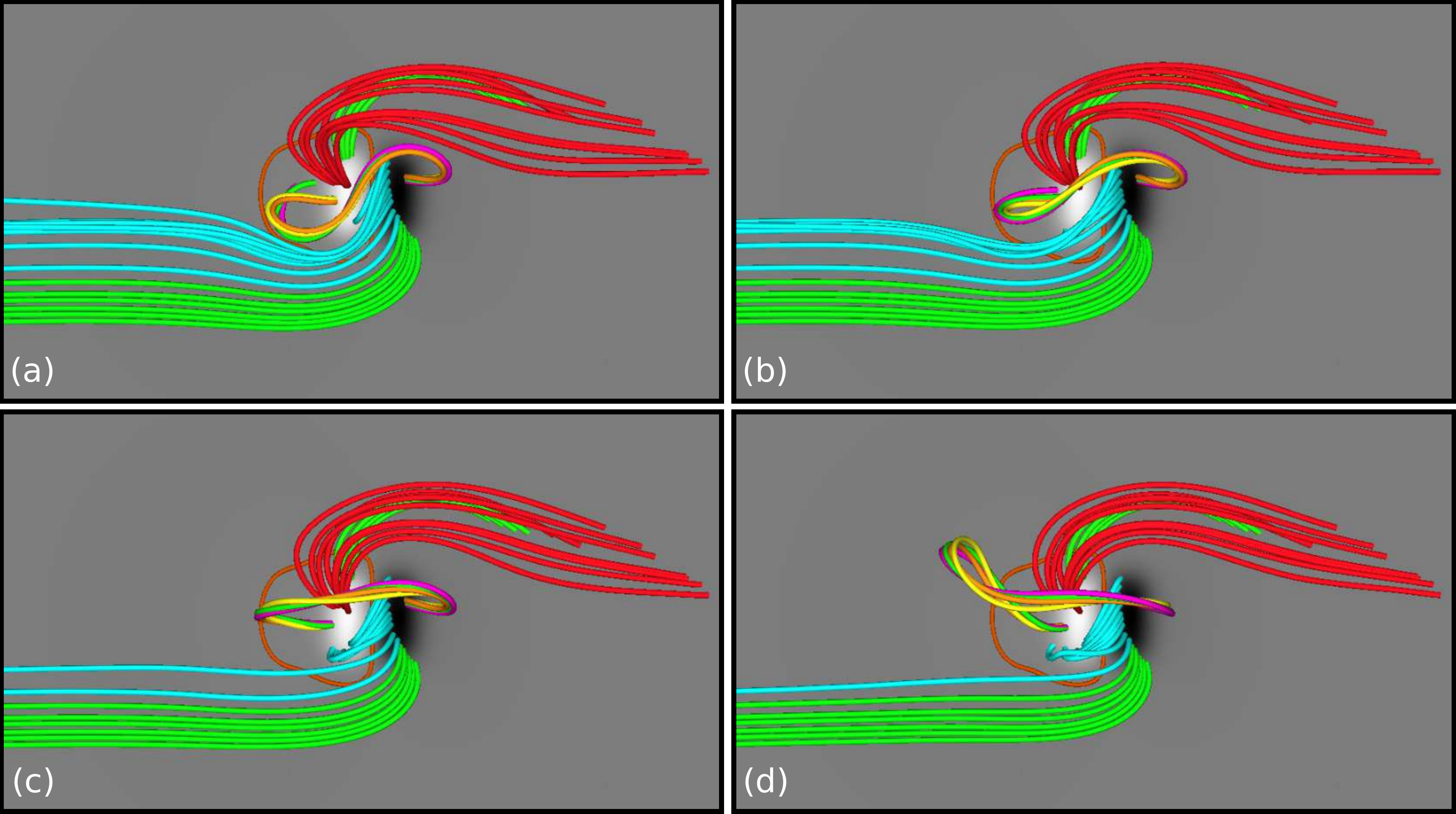}
  \caption{Field lines (yellow, orange, and magenta) near the axis of the flux rope showing the development of writhe. Red, green, and cyan field lines and surface shading are as in Fig.\ \ref{fig:flstop}. An animation is available online. (a) $t \approx 48$\,min $20$\,s. (b) $t = 49$\,min $30$\,s. (c) $t \approx 50$\,min $40$\,s. (d) $t \approx 51$\,min $50$\,s.}
  \label{fig:core}
\end{figure*}

Figure \ref{fig:flsangle} shows the same evolution from a different point of view, more clearly demonstrating the transfer of twist when the flux rope reconnects. Note the height attained by the erupting flux rope as it approaches open-closed reconnection (Fig.\ \ref{fig:flsangle}(b)). The apex is considerably higher than the breakout current sheet (which resides near the surface), suggesting that the flux rope experiences internal forces that push it upward into the strong, uniform background field. To investigate whether the flux rope experiences an ideal kink-like instability in the late stages of breakout, we studied the field-line evolution in detail around this time. Figure \ref{fig:core} shows selected field lines close to the axis of the flux rope during the eruption. The axis is sigmoidal initially, but then rapidly rotates, straightening out and then bending back on itself into an inverse-$\gamma$ shape. This is the classic evolution of a kink-unstable flux rope, which converts twist within the rope into a writhe of the rope axis \citep{Torok2005}. The ideal nature of the instability is further confirmed by the fact that the footpoints of the axis field lines do not change during this evolution (see also the online animation).

\begin{figure}
	\includegraphics[width=\columnwidth]{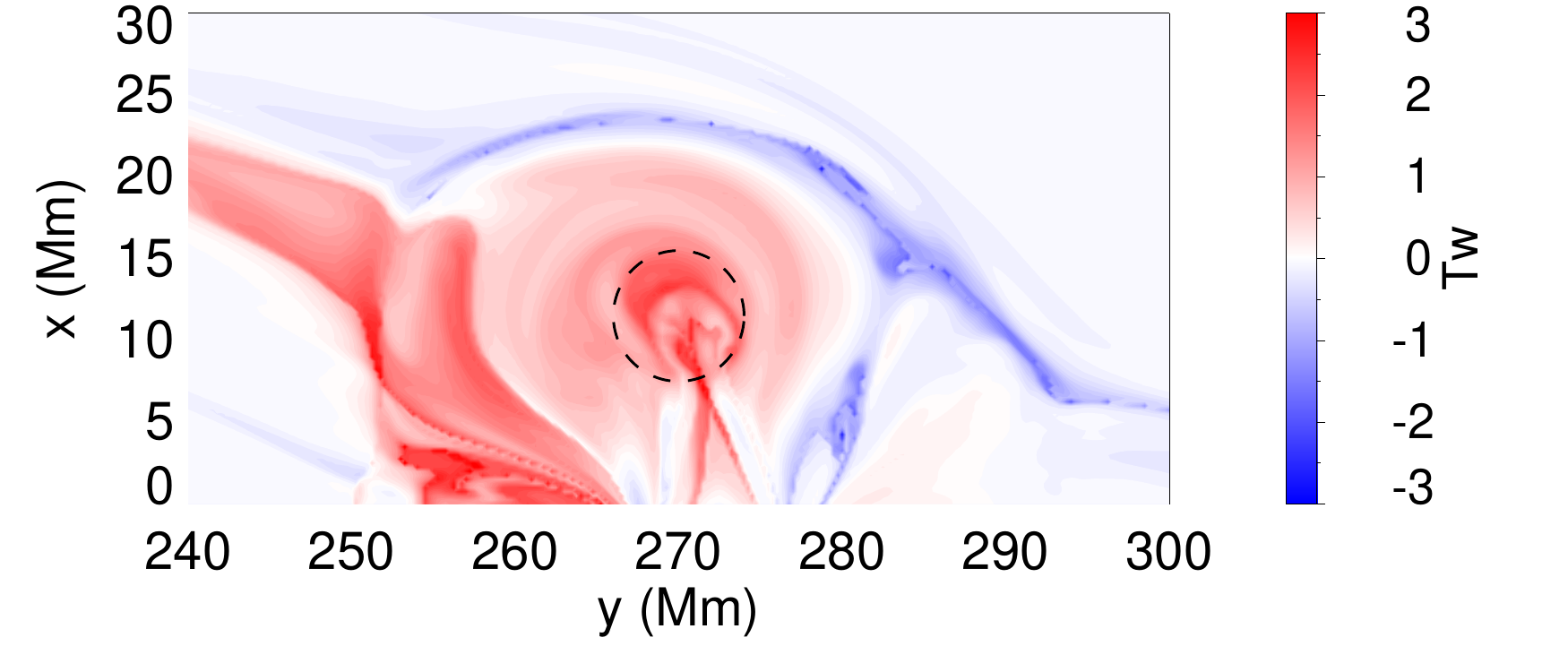}
  \caption{$T_w$ calculated in a plane crossing the flux rope at $t \approx 48$\,min $20$\,s. The black dashed lines show the circle within which the average is calculated.}
  \label{fig:twist}
\end{figure}

{Additional evidence that this is indeed a kink instability can be obtained from the number of field-line turns within the flux rope just prior to the onset of fast dynamical evolution. Dedicated simulation studies have revealed that the critical number of turns required for the kink instability varies between $1.25$ and $1.75$ turns, depending upon the properties of the flux rope and its strapping field \citep[e.g.][]{Torok2003,Torok2004}. We estimate the number of turns present in our flux rope by calculating the twist number, $T_w$, defined as
\begin{align}
T_{w} = \int_{s}{\frac{(\nabla\times\mathbf{B})\cdot \mathbf{B}}{4\pi B^{2}}ds} = \int_{s}{\frac{\mu_{0} J_{\parallel}}{4\pi |B|}ds}
\end{align}
where the integral is calculated along each field line \citep{Berger2006}. $T_w$ measures the average number of turns locally about a given field line. Figure \ref{fig:twist} shows $T_w$ within a plane ($z=0$) bisecting the flux rope at $t \approx 48$\,min $20$\,s, just prior to instability onset. The approximate position where the flux rope crosses this plane is shown by the dashed black circle. There is significant variation of $T_w$ within the flux rope, with values varying between $\approx 1$ and $\approx 3$ turns, reflecting the fact that the flux rope forms dynamically in our simulation. Taking an average of $T_w$ within the circular region yields an average twist of $\approx 1.55 \pm 0.1$ turns. The error bars were obtained by calculating the average using a few different radius values. This average value lies within the expected range of critical twists, strongly supporting our conclusion that the evolution is due to kink instability.}

{All of the above suggest that, in the late stages of breakout, the closed-closed reconnection occurring below the flux rope generates enough twist that the flux rope becomes unstable to kinking. The conversion of this twist into writhe due to the kink instability causes the flux rope to rotate and expand upward more quickly. With the onset of the kink, the breakout reconnection rapidly accelerates until the flux rope reaches the overlying null, breaks through, and begins to reconnect with the external field.} The kink instability, therefore, acts to super-charge the breakout process in its final stage. 

\begin{figure}
	\includegraphics[width=\columnwidth]{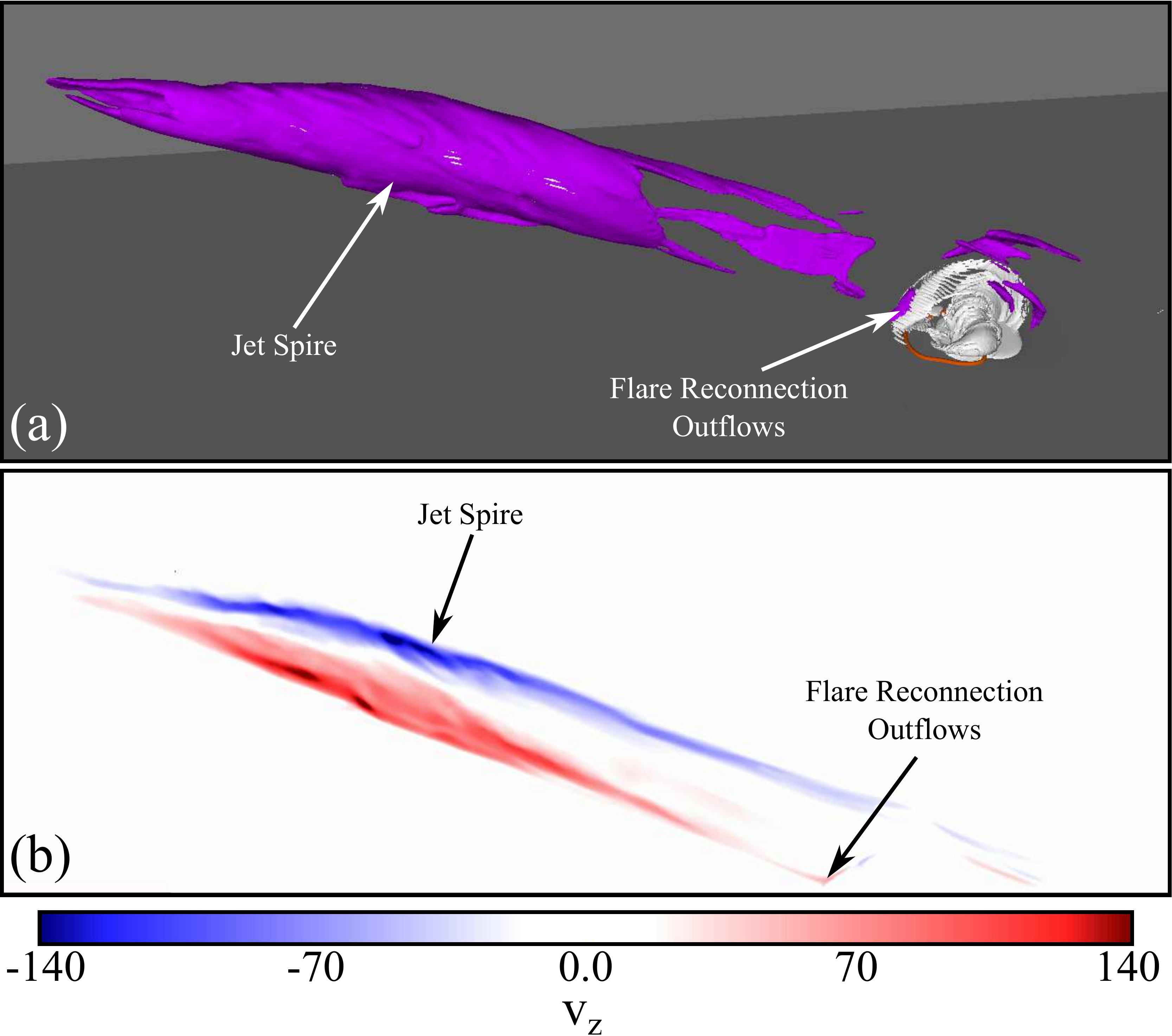}
  \caption{Jet morphology at $t \approx 59$\,min. (a) Isosurfaces of current density (white, $|J| = 3.6\times10^{-3}$ A m$^{-2}$) and plasma velocity magnitude (purple, $|v| = 141$ km s$^{-1}$). Surface shading is as in Fig.\ \ref{fig:flstop}. (b) Vertical cut showing the rotational component of the plasma velocity ($v_{z}$) within the jet spire.}
  \label{fig:iso}
\end{figure}

This accelerating evolution culminates in an untwisting helical jet generated by a combination of twist propagation {in the form of non-linear Alfv\'{e}n waves} and open-closed reconnection outflows from the flare current sheet. In this case, the former dominates, producing a strongly rotating jet spire. Figure \ref{fig:iso}(a) shows an isosurface of velocity magnitude in the broad spire {above the less dominant outflows from the curved flare current layer}; the rotational component of the vector velocity is shown in Figure \ref{fig:iso}(b). Figure \ref{fig:current}(b) shows a close-up view of the current structures in the jet base at this time. The flare current sheet (FCS) sits on the open-closed boundary, passing through the null point (near the surface) and arching over the embedded bipole. The black arrows indicate the direction of reconnection inflows and outflows within the sheet, showing how the flare reconnection acts to reverse the reconnection of the breakout phase (Fig.\ \ref{fig:current}(a)) and to return the field regions back towards their original states. That is, the flare reconnection re-closes cyan field lines and re-opens red ones. Note, however, that the flare reconnection achieves this flux transfer in a much shorter time than the breakout reconnection prior to the jet. 

%t = 820, 840, 860, 880

%t = 820

%|J| = 0.3, |v| = 0.1, t = 100

\begin{figure}
	\includegraphics[width=\columnwidth]{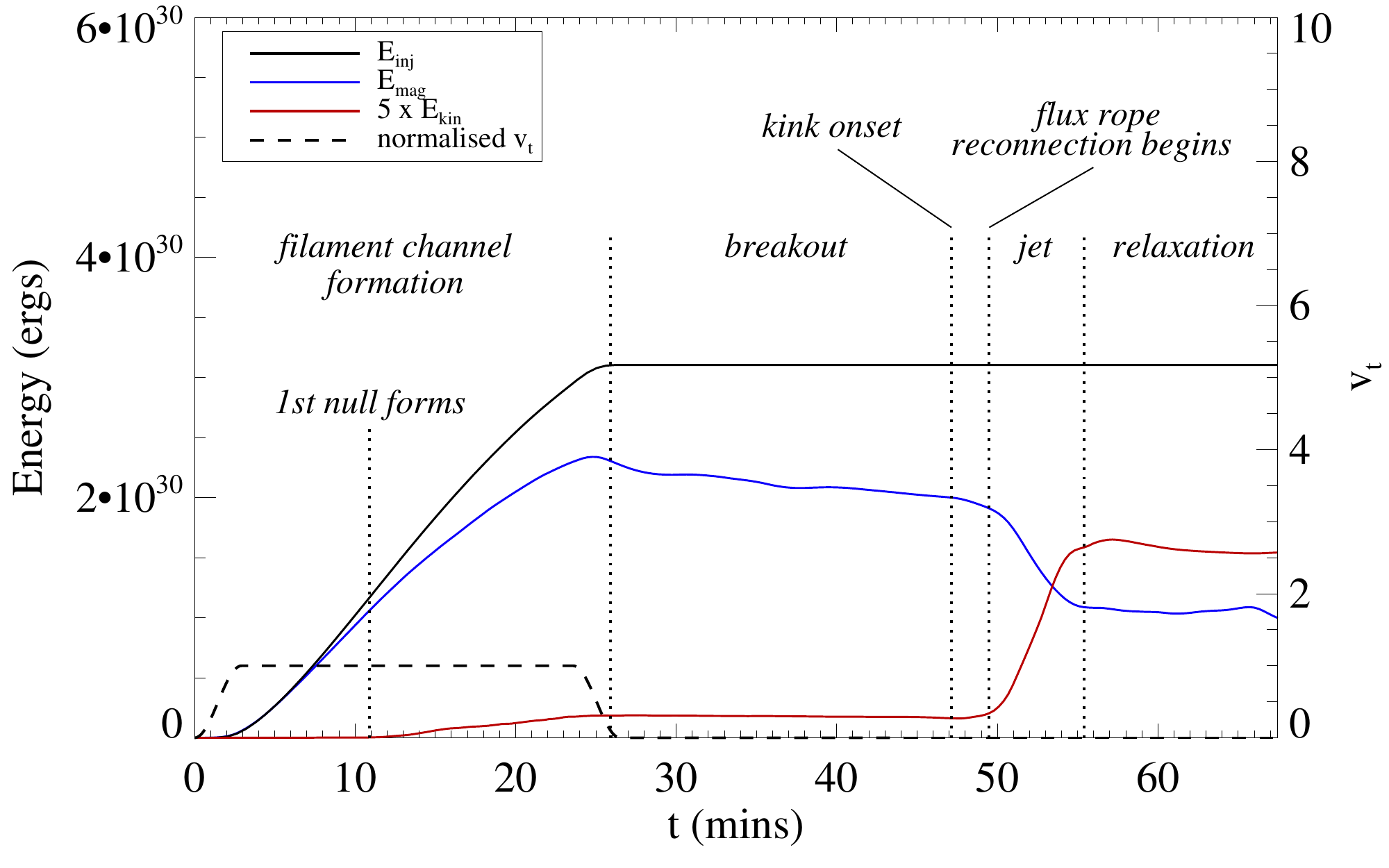}
  \caption{Volume-integrated free magnetic (blue, $E_{mag}$) and kinetic (red, $E_{kin}$) energies. The black curve shows the cumulative integral of the Poynting flux across the bottom boundary, showing the energy injected by the surface driving over time ($E_{inj}$). Shown in dashed black is the time profile of the driver ($v_{t}$, normalised to one). }
  \label{fig:energies}
\end{figure}

\subsection{Energies}
Figure \ref{fig:energies} shows the volume-integrated free magnetic (blue) and kinetic (red) energies with the different stages of the jet simulation highlighted. Also shown are the cumulative Poynting flux injected by the surface driving (solid black) and the time profile of the driving (dashed black). The bald patch is superseded by a null point early in the filament-channel formation phase, after which the quasi-steady reconnection outflows slowly increase the total kinetic energy ($11$\,min $< t < 26$\,min). The deviation of the blue and black curves shows that this quasi-steady reconnection releases some free energy during this phase. Once the driving stops, the slow breakout evolution takes over, releasing a small amount of free energy ($26$\,min $< t < 49$\,min). {The total kinetic energy remains nearly constant during this time, showing that the kinetic energy produced by the breakout flows is small and approximately matches the gradual numerical viscous dissipation of the previously generated flows as they propagate away. The free energy that is released, but not converted to kinetic energy, is lost from the system through numerical dissipation within the reconnecting current layers. In a simulation with more comprehensive, and far more computationally demanding, coronal thermodynamics, this additional energy would be converted into heat and then convected, conducted, or radiated away.} The onset of the kink motion coincides with a slight drop in free energy and slight increase in kinetic energy at the end of the breakout phase ($47$\,min $< t < 49$\,min). Significant free-energy release starts when the flux rope begins to reconnect, converting its magnetic twist and writhe to kinetic energy of untwisting motions within the jet ($49$\,min $< t < 55$\,min). The kink evolution of the flux-rope axis (Fig.\ \ref{fig:core}) lasts until the end of the jet phase, as these core field lines are some of the last to be reconnected. Negligible free energy is released following the jet, as the flare reconnection dies away and the jet base relaxes towards a new equilibrium state ($t > 55$\,min).

\begin{figure}
	\includegraphics[width=\columnwidth]{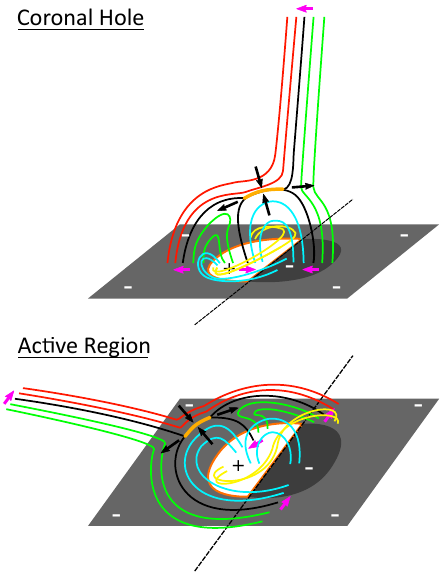}
  \caption{Schematic comparing the breakout process in the active-region and coronal-hole models. Black arrows show the direction of reconnection inflows and outflows at each breakout current sheet. Pink arrows show the direction in which the spines and fan plane move in response to this reconnection. The PIL is shown in dark orange; the dashed line is aligned with the PIL at the centre of the embedded bipole.}
  \label{fig:schematic}
\end{figure}

\begin{figure}
	\includegraphics[width=\columnwidth]{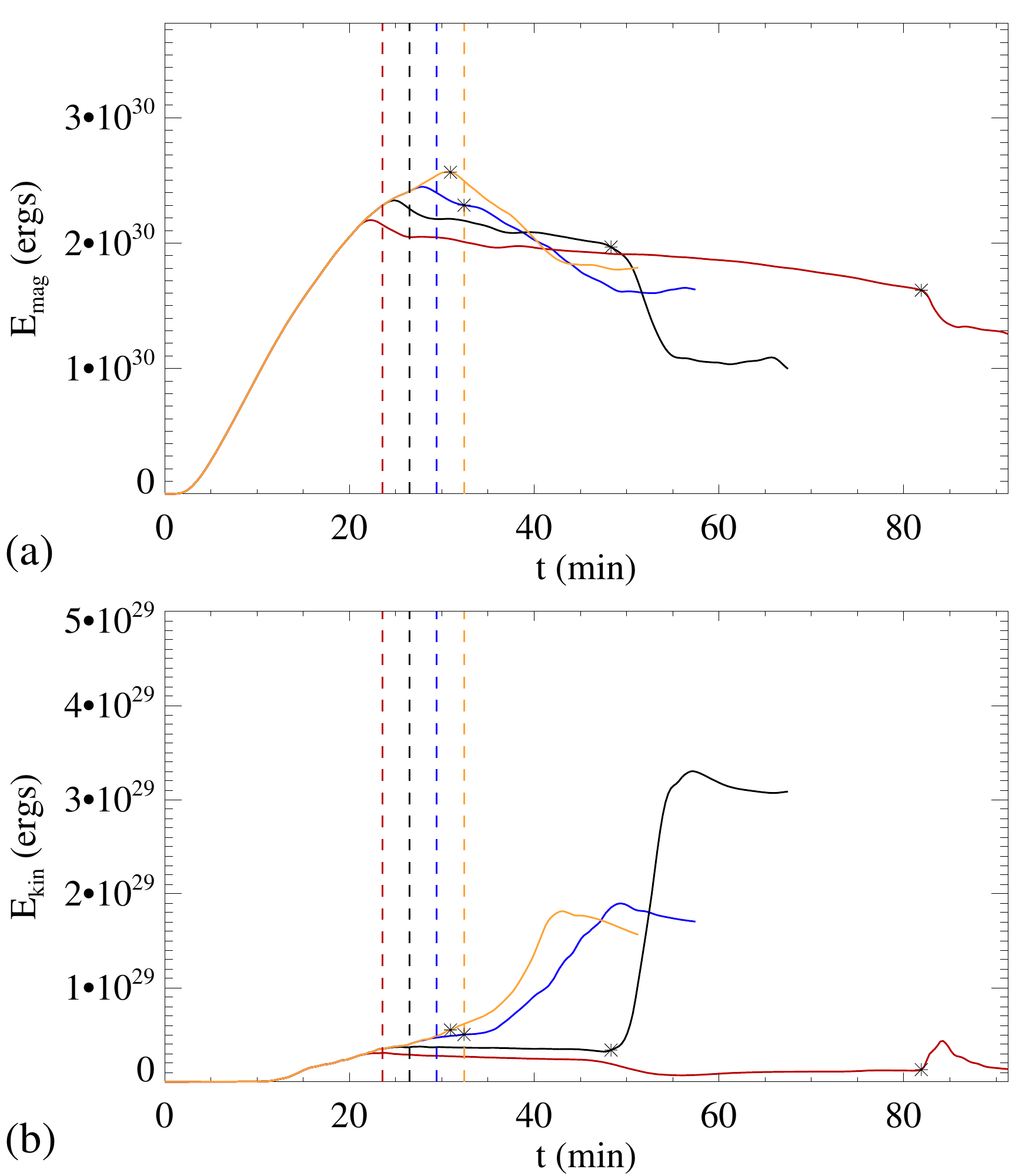}
  \caption{Volume-integrated (a) magnetic and (b) kinetic energies for four simulations with different driving periods. Vertical dashed lines indicate when the driving ceases in each simulation. Red: $t \approx 23$\,min $30$\,s. Black: $t \approx 26$\,min $30$\,s. Blue: $t \approx 29$\,min $30$\,s. Yellow: $t \approx 32$\,min $30$\,s. Asterisks mark approximately when the jet begins in each simulation. Black curves represent the main simulation (Fig.\ \ref{fig:energies}).}
  \label{fig:comp}
\end{figure}

\begin{figure*}
	\includegraphics[width=2\columnwidth]{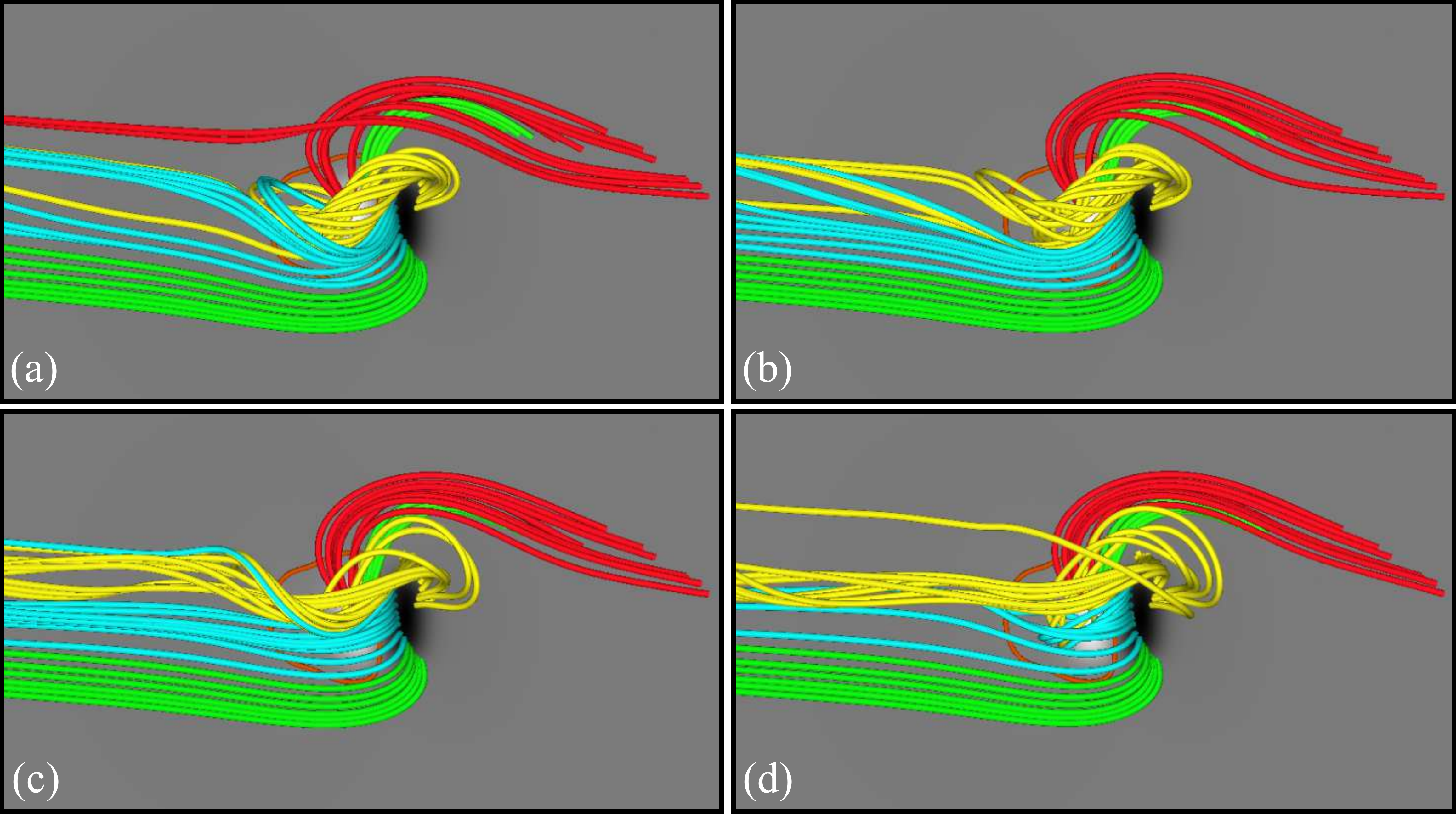}
  \caption{Top view of field lines showing the reconnection of the flux rope when the driving is maintained until $t \approx 29$\,min $30$\,s. (a) $t \approx 32$\,min $25$\,s. (b) $t \approx 35$\,min. $20$\,s (c) $t \approx 38$\,min $18$\,s. (d) $t \approx 41$\,min $15$\,s. Surface shading as in Fig.\ \ref{fig:setup}. The PIL is shown in orange.}
  \label{fig:fls450}
\end{figure*}

\section{Discussion}
\label{sec:discussion}

\subsection{Comparison with Previous Models}
In the early stages of the simulation, a current sheet forms at the bald patch in response to the expansion of the closed field beneath the separatrix. As this current sheet forms on the simulation boundary, it is important to understand the impact of the boundary conditions on this aspect of our results. The line-tying at the boundary pins the magnetic field lines at the bald patch to the surface, approximately replicating the effect of the rapid increase in thermal pressure at the chromosphere and photosphere. Line-tying allows us to model large coronal domains, making this 3D jet simulation practical, without having to resolve the lower layers of the solar atmosphere. However, this simplification affects the dynamics within and just above those layers. For example, \citet{Karpen1990} demonstrated that when these layers are explicitly treated and a bald-patch separatrix is gently driven, the field lines are able to rise sufficiently that no bald-patch current sheet forms. In our case, the entire closed-field region is subject to a systematic expansion that affects not just field lines touching the bald patch, but also those that would penetrate deep into the lower atmosphere were those layers included. These more deeply rooted field lines would be unable to move freely, so it is reasonable to expect that a current sheet would form in a manner similar to that occurring in our simulation \citep[see also][]{Titov1993}. In flux-emergence simulations with pre-existing coronal field, similar current sheets are observed to form dynamically at bald patches, created by ``U-loops'' that dip below the photosphere \citep[e.g.,][]{Cheung2010}. 

The topology change within the bald-patch current sheet is also affected by the line-tying. Were a sub-surface layer included, the null point we identified as the surface null would not have formed; only the upper and lower nulls, which form away from the boundary, would have been created. Thus, the annihilation of the lower null with the surface null is also a result of the boundary condition. Only the upper null facilitates the subsequent breakout reconnection and ultimate generation of the jet, however. The lower null would be expected to remain near the surface, well away from the jet dynamics. In fact, the orientation of the spine of the lower null must be parallel to the direction of current, so that it is of spiral type \citep[see, e.g.,][for details]{Wyper2014a,Wyper2014b}. The spiral field lines near this null would embed it in the dense, lower layers of the atmosphere. We conducted a test simulation of the early phase of our jet simulation that included a chromosphere, and we confirmed that the lower null indeed remains in the low atmosphere. The simulation was impractical to run beyond that point, however. Those results will form the basis of a separate, forthcoming publication. We note that similar results have been found in recent flux-emergence numerical experiments \citep{Leake2018}. Therefore, although some low-lying structures are not captured entirely faithfully in our simulation reported here, they have no significant effect on the high-lying structures -- filament channel and upper null -- that form the jet once the external reconnection is initiated. 

The current sheet that forms at the upper null supports breakout reconnection that slowly removes strapping flux from the side of the flux rope. This differs from the breakout evolution in our coronal-hole model, in which strapping flux is removed from above the centre of the rising flux rope. A schematic diagram summarising the breakout evolution in each model is shown in Figure \ref{fig:schematic}. The PIL is shown in dark orange, and dashed lines show the orientation of the section of the PIL at the centre of the bipole, along which the filament channel forms. Cyan field lines show the two regions of strapping flux, one across the middle and the other along the side of the flux rope/filament channel. The high position of the null point in the coronal-hole setting (top) leads to reconnection of the central strapping flux, allowing the flux rope to rise upwards. In this case, the footpoints of the inner spine and fan plane move perpendicular to the filament channel (pink arrows). {In contrast, in the active-region setting (bottom) the field is highly inclined and approximately aligned with the central strapping field. The low position of the null, once formed, puts it in proximity of the side of the flux rope.} Removal of this side flux leads to a sideways expansion and rotation of the flux rope, yet a similar feedback between breakout reconnection and flux-rope movement occurs here, as well. In this case, the movement of the footpoints of the inner spine and fan plane is essentially parallel to the filament channel. 

The field is inclined $70^{\circ}$ relative to the vertical in this simulation. In \citet{Wyper2018}, we tested three different field inclinations and found that breakout reconnection of the coronal-hole type occurred in each. The steepest field inclination that we tested was $22^{\circ}$. A transition between the coronal-hole and active-region breakout evolutions evidently occurs in the range $[22^{\circ},70^{\circ}]$. Since breakout reconnection in our new simulation involves sheared field lines, however, the transition may depend somewhat upon the detailed field structure in the filament channel.

We found further that the flux rope develops writhe just prior to and during the jet, demonstrating the likely onset of a kink instability within the flux rope \citep{Torok2005}. Prior to the onset of the instability, the system evolves slowly with breakout and tether-cutting reconnection gradually removing strapping flux from around the flux rope and advancing it toward the onset of open-closed reconnection. The onset of the kink instability rapidly speeds up the final stage of this process. {That is, once the kink is triggered, the flux-rope rotation rapidly increases compared with its slow evolution during the breakout phase. This rapid rotation and expansion then drive faster breakout reconnection.} However, little energy is released until the flux rope begins to reconnect onto open field and the jet is launched. At that point, rapid open-closed reconnection of the flux rope occurs and ideal untwisting of the newly reconnected open field lines drives the fast jet flows. In our coronal-hole simulations, the onset of fast dynamics occurred essentially when the flux rope began to reconnect. {We saw little obvious evidence of a rapid increase in the flux-rope rise or kink in those simulations,} although it is possible that the flux ropes that developed may have accumulated enough turns to kink just before they reconnected, and any writhe simply had no time to develop. 

To better understand the role of the kink instability in this setup, we conducted three additional simulations where the driving was halted at earlier or later times. The energy curves for all four simulations are shown in Figure \ref{fig:comp}, where vertical dashed lines indicate when the driving was halted in each case. These tests revealed that if the driving was maintained, a jet was generated {without a flux-rope kink.} {An example of this is shown in Figure \ref{fig:fls450} for the simulation corresponding to the blue curve in Figure \ref{fig:comp}; the case corresponding to the yellow curve behaves similarly. The flux rope is in a rotated position just prior to reconnection, similar to that in the kink unstable case, Figure \ref{fig:fls450}(a) (compare with Figure \ref{fig:flstop}(b)). In the kink-unstable case, the flux rope then rises and rotates as the instability creates writhe in the flux rope, Figures \ref{fig:flstop}(c) and \ref{fig:flsangle}(c). On the other hand, when the driving is maintained, the external current sheet reaches the flux rope before any kink instability develops, and it reconnects the flux rope directly, Figure \ref{fig:fls450}(b)-(d). The result is a less impulsive release of the twist within the flux rope over a longer time, as shown by the blue energy curves in Figure \ref{fig:comp}.} 

{These results are rather analogous to the jets driven by flux emergence, in that the imposed flows directly drive the flux rope to reconnect with the external field. It is plausible that imposing slower surface motions would allow the flux rope to acquire sufficient twist to induce kink instability before it encounters the null point and external magnetic flux, as occurred in our baseline simulation. The free energy stored in our maintained-driving cases is, in fact, greater than that in our kink-unstable case, as shown in Figure \ref{fig:comp}. However, demonstrating the conjectured kink-instability onset would require prohibitively long numerical calculations. }

In a test where the driving was halted sooner (injecting less free energy, red curves), the breakout phase was longer, but eventually the flux rope kinked, triggering a small jet. This delay may follow from the extra time needed to develop enough turns in the flux rope to trigger the instability. There is a drop in kinetic energy in this simulation around $t = 48$\,min, due to the earliest reconnection outflows and waves propagating out of the simulation domain. {Our limited parameter study therefore suggests that, when surface motions do not dominate the evolution up to the time of jet onset, a kink instability of the flux rope may be necessary to trigger jets in the highly inclined ambient fields at the edges of active regions where bald-patch or near-surface-null topologies are prevalent.} The breakout reconnection acts to generate tether-cutting reconnection, which slowly builds up the flux rope and its twist in this case. A fuller parameter study is needed to explore these conclusions comprehensively, however. 

{Although our simulations are energised by surface motions, as discussed in \citet{Wyper2018} the breakout mechanism is rather generic and is expected to apply to other processes, such as flux emergence or flux cancellation, for generating excess free magnetic energy within the closed field beneath the separatrix. Indeed, results similar to our coronal-hole jet model have been observed in long-duration flux-emergence experiments with oppositely aligned overlying field \citep[e.g.\ ][]{Archontis2013,Moreno-Insertis2013,Fang2014}. Although these investigations do not go into great detail about the triggering mechanism, they exhibit a similar combination of external and internal (breakout and tether-cutting) reconnection leading to a transfer of twist and the formation of helical jets. It seems likely, therefore, that our present result of a slow breakout evolution coupled to a fast ideal kinking should also be expected to occur where the free energy is injected through some other mechanism.}

\subsection{Comparison with Observations}
Our simulation agrees well with a number of features of jets observed at the periphery of active regions. Qualitatively, our model explains how mini-filaments observed at the edge of some active regions erupt to produce helical jets \citep[e.g.,][]{Yan2012,Hong2016,Sterling2016,Zhu2017}. The spires of these jets usually contain both cool and hot components, consistent with the ejection of cool filament plasma mixed with plasma heated by reconnection \citep[e.g.,][]{Mulay2017}. Although our model does not include the cool filament plasma directly, it predicts a similar behaviour based on the evolution of the associated magnetic structures. Helical active-region jets tend to have multiple flare kernels and bright points. In particular, surface brightenings and EUV loops are observed to form some distance away from the original location of the mini-filament, towards the active region. The field-line evolution shown in Figure \ref{fig:current}(b) during the impulsive flare-reconnection phase gives a simple explanation for this: the strongly inclined ambient field implies that the footpoints of the overlying flare-reconnected field lines lie far from the bipole, in the direction of the active region. Additionally, bright loops that arch over the bipole are often observed to persist in the minutes after the jet is launched \citep[e.g.,][]{Zhu2017}. The arching shape of the flare current sheet in Figure \ref{fig:current}(b) provides a plausible explanation for these loops as forming due to reconnection {\it{parallel}} to the filament channel.

Quantitatively, our simulation also compares well with jets observed on the periphery of active regions. Typical observed jet speeds and lifetimes range between $87 - 532$\,km\,s$^{-1}$ and $5 - 39$\,mins, respectively \citep{Mulay2016}. The helical jet phase of our simulation lasted $\approx 5$\,mins (Fig.\ \ref{fig:energies}) and exhibited peak plasma flow speeds of $\approx 140$\,km\,s$^{-1}$ (Fig.\ \ref{fig:iso}), falling near the lower end of both observed ranges. However, note that different choices for our scaling parameters would yield longer-lived, faster jets. {For instance, $L_{s} = 13.5$\,Mm, $\rho_{s} = 2\times 10^{-14}$\,g\,cm$^{-3}$, and $B_{s} = 28.5$\,G give a jet phase duration of $\approx 10$\,mins with a peak plasma flow speed of $\approx 200$\,km\,s$^{-1}$.}

Topologically, magnetic-field extrapolations of jet source regions have revealed a mixture of bald-patch and null-point topologies \citep[e.g.,][]{Schmieder2013,Mandrini2014,Zhu2017,Chandra2017}. When a bald patch is identified, it is sometimes presumed to persist throughout the lifetime of the jet. Our simulation shows that when dynamics are included, the bald patch is quickly replaced by a null point, which forms dynamically in the low solar atmosphere before rising higher into the corona. Precisely identifying the layer in which the null forms (photosphere, chromosphere, or transition region) requires a more comprehensive simulation with these layers included, a topic for future work. Significant reconnection outflows occur in our simulation only after the null has formed and risen into the corona. Therefore, wherever there are observed jet outflows, it should be suspected that a low coronal null is present, even if field extrapolations suggest otherwise. The presence of a null point during the jet is important, as 3D nulls are potential sites of particle acceleration \citep[e.g.,][]{Dalla2006,Stanier2012,Baumann2013} and helical active-region jets are thought to be sources of some impulsive SEP events \citep[e.g.,][]{Bucik2018}. 

Finally, many active-region-periphery jets have a recurrent nature \citep[e.g.,][]{Liu2016,Chandra2017}. Multiple jets can originate from the same region in succession, if energy is explosively released episodically while continually being injected slowly into the closed field beneath the separatrix by sustained surface motions, flux emergence, and/or flux cancellation. In our simulation, we used surface motions as a numerically convenient way to introduce the free energy. The model could easily be extended to simulate homologous behaviour, simply by maintaining the surface driving motions or periodically switching them on and off repeatedly \citep{Pariat2010}. The filament channel then would reform before eventually erupting to drive another helical jet. 

\section{Conclusions}
\label{sec:summary}
In this work we extended our model for coronal-hole jets \citep{Wyper2017,Wyper2018} to a configuration typical of jets from the periphery of active regions. Our main findings are as follows:
\begin{itemize}
\item Although we start with a bald-patch configuration, early in the evolution a coronal null point forms that facilitates breakout reconnection. The breakout reconnection removes strapping field constraining the end of the filament flux rope, rather than its centre as in the coronal-hole model.
\item Rather than rising, the flux rope that forms expands laterally and rotates towards the null. Simultaneously, {closed-closed (tether-cutting) reconnection} beneath the flux rope increases the twist within it.
\item Rapid evolution, but not rapid energy release, is initiated by the onset of a kink instability of the flux rope, which quickly rises and rotates as it converts twist into writhe. This accelerates the breakout reconnection until the flux rope reaches the breakout current sheet, whereupon it reconnects onto open field, resulting in explosive energy release. 
\item As in the coronal-hole model, reconnection of the flux rope launches nonlinear Alfv\'{e}n waves and fast reconnection outflows from the flare current sheet behind the flux rope, producing a broad, helical jet spire. 
\end{itemize}
Our results show that energetic jets from the periphery of active regions are similar in nature to those from coronal holes. However, the high field inclination and low coronal-null position combine to store comparatively more free energy than in coronal holes, where null reconnection occurs more readily as the filament channel forms. In these ARP jets, the explosive jet onset clearly is driven by coupling between the ideal instability of the flux rope and the non-ideal breakout reconnection.

The same physical behavior was observed in previous coronal-hole jet simulations that had no filament channel \citep[e.g.,][]{Pariat2009,Pariat2010,Pariat2015,Pariat2016,Karpen2017}. In those cases, the photospheric motions injected into the corona a large-scale twist, whose width was comparable to that of the whole closed-flux region of the embedded bipole. This global twist built up until the closed flux underwent a global, kink-like instability. The closed flux expanded upward until it buckled, strongly compressing the flux toward the null point and inducing fast reconnection there. Prior to onset of the instability, breakout reconnection clearly occurred, but it was slow and produced only a small energy release. Conversely, the ideal instability alone also produced negligible energy release, as demonstrated by a simulation of the system with a flux-preserving Lagrangian code that enforced purely ideal evolution \citep{Rachmeler2010}.  As with the kink-driven jets of this paper, the explosive energy release in those tilt-driven jets clearly was due to the coupling between breakout reconnection and the ideal instability.

These results suggest that such ideal-instability/breakout-reconnection coupling may be responsible for jet onset in all of the models that we have studied. If so, the question arises as to the nature of the ideal instability in our previous jet studies with filament channels, in which we found no clear evidence of kink instability \citep{Wyper2017,Wyper2018}. Rather, we found that the onset of flare reconnection within the filament-channel flux system induced an accelerating rise of the flux rope toward the overlying null. Perhaps the flare reconnection accumulates sufficient twist and net electric current in the flux rope to initiate a torus instability \citep{Kliem2006}; in that case, breakout reconnection across the null-point current sheet effectively provides a large decay index in the strapping field and, hence, instability of the flux rope. Alternatively, the system may have reached a configuration in which the overlying field was no longer able to confine the flux rope in equilibrium, initiating a slow but inexorable upward expansion of the flux rope. In either case, the coupling between breakout reconnection and ideal instability (or loss of equilibrium) may well be critical for ultimately achieving fast energy release. Conclusive tests of these ideas require an investigation of the flux-rope evolution under strictly ideal motion \citep[e.g.,][]{Rachmeler2010}.

If the above conjectures are correct, then {the general mechanism for generating eruptive jets may be a resistive instability coupling breakout reconnection to some underlying ideal instability.} The specific ideal instability may depend upon the particular characteristics of the system, but the basic physics are the same. Further modelling and observational work are needed to confirm or refute this explanation for the fascinating phenomenon of solar coronal jets.

\section*{Acknowledgments}
This work was supported through a Fellowship award to PFW by the Royal Astronomical Society and grant awards to CRD and SKA by NASA's H-ISFM, H-LWS, and H-SR programs. Computer resources for the numerical calculations were provided to CRD by NASA's High-End Computing program at the NASA Center for Climate Simulation. 

%%%%%%%%%%%%%%%%%%%%%%%%%%%%%%%%%%%%%%%%%%%%%%%%%%



%\bibliographystyle{mnras}
%\bibliography{biblionew} % if your bibtex file is called example.bib

%\bsp	% typesetting comment
%\label{lastpage}
%\end{document}

%\begin{thebibliography}{99}

%%%%%%%%%%%%%%%%%%%%%%%%%%%%%%%%%%%%%%%%%%%%%%%%%%

%%%%%%%%%%%%%%%%% APPENDICES %%%%%%%%%%%%%%%%%%%%%
%\appendix

%%%%%%%%%%%%%%%%%%%%%%%%%%%%%%%%%%%%%%%%%%%%%%%%%%

% Don't change these lines
\bsp	% typesetting comment
\label{lastpage}
\end{document}